----------------------------------------------------------------

\documentstyle[12pt]{article}
\title{A New Solution of the Yang-Baxter Equation\\
Related to the Adjoint Representation of $U_{q}B_{2}$}

\author{Zhong-Qi Ma\\
{\small CCAST(World Laboratory), P.O.Box 8730, Beijing 100080
and} \\[-2mm]
{\small Institute of High Energy Physics, P.O.Box 918(4),
Beijing 100039, P. R. of China}\\
and\\
An-Ying Dai\\
{\small Beijing Institute of Technology, Beijing 100081,
P. R. of China}}

\date{}
\begin{document}
\maketitle

\vspace{20mm}

\begin{abstract}
A new solution of the Yang-Baxter equation, that is related to
the adjoint representation of the quantum enveloping algebra
$U_{q}B_{2}$, is obtained by fusion formulas from a non-standard
solution.
\end{abstract}

\newpage
\noindent
{\bf 1. INTRODUCTION}

\vspace{3mm}
There are three typical methods [1] for finding the
trigonometric solutions of the Yang-Baxter equation [2]. The
main one is based on Jimbo's theorem [3,4]. The necessary
condition for using this method is existence of the quantum
generator $e_{0}$, corresponding to the negative lowest root.
The second method for finding solutions is so-called
Yang-Baxterization, namely to embed appropriately the spectral
parameter $x$ into a solution $\breve{R}_{q}$ of the simple
Yang-Baxter equation such that $\breve{R}_{q}(x)$ satisfies the
Yang-Baxter equation. This method is useful for the cases where
the spectrum-independent solution $\breve{R}_{q}$ has only two
or three different eigenvalues [5,1]. The third method is the
fusion formulas [6,1] where an appropriate project operator is
needed.

Unfortunately, firstly, the explicit form of $e_{0}$, that
satisfies the quantum algebraic relations, does not exist for
the adjoint representation of any quantum enveloping algebra
$U_{q}{\cal G}$, except for $U_{q}A_{\ell}$. Secondly, the
spectrum-independent solution $\breve{R}_{q}$ for the adjoint
representation usually has much more different eigenvalues than
three. For instance, in the simplest case, the solution
$\breve{R}_{q}$ for the adjoint representation of $U_{q}B_{2}$
has six different eigenvalues. At last, from the solution
$\breve{R}_{q}(x)$ related to the minimal representation,
obtained based on Jimbo's theorem, the needed project operator
for the fusion formulas does not exist for this case. It is the
reason why no solution of the Yang-Baxter equation related
to the adjoint representation of $U_{q}{\cal G}$, except for
$U_{q}A_{\ell}$, was found up to now.

On the other hand, by Yang-Baxterization when $\breve{R}_{q}$
has three different eigenvalues, there is an additional
solution, so-called non-standard one, that happens to provide
the needed project operator for the fusion formulas. In this
way we are able to compute the solution related to the adjoint
representations of $U_{q}B_{\ell}$, $U_{q}C_{\ell}$ and
$U_{q}D_{\ell}$. In order to realize this idea, in this paper
we compute explicitly the simplest example of those cases: the
trigonometric and rational solutions of the Yang-Baxter equation
related to the adjoint representation of $U_{q}B_{2}$, that is
equivalent to $U_{q}C_{2}$. The rest of solutions can be computed
straightforwardly, but more complicatedly.

This paper is organized as follows. In Sec. 2, we show that the
explicit form of $e_{0}$ matrix for the adjoint representation
of $U_{q}B_{2}$, that satisfies the quantum algebraic relations,
does not exist. In order to use the fusion formulas, we have to
compute firstly the solution $\breve{R}_{q}(x)$ of the
Yang-Baxter equation related to the minimal representation in
Sec. 3. From it we obtain the project operator
$\breve{R}_{q}(q^{-4})$ that maps the direct product spaces
$V_{(1~0)}\otimes V_{(1~0)}$ onto the representation space
$V_{{\rm adj}}=V_{(0~2)}$ of the adjoint representation, where
$V_{(1~0)}$ is the representation spaces of the minimal
representation (1~0). In Sec. 4 we sketch the proof for the
fusion formulas. The explicit form of $\breve{R}_{q}^{{\rm
adj}}(x)$ is computed in Sec. 5 in terms of the quantum
Clebsch-Gordan coefficients for the coproduct in the direct
product of two representation spaces of the adjoint
representation. The corresponding rational solution of the
Yang-Baxter equation is obtained in Sec. 6 by a standard limit
process [1].

\vspace{15mm}
\noindent
{\bf 2. Non-Existence of $e_{0}$ Matrix}

\vspace{3mm}
The Cartan matrix for the algebra $B_{2}$ is
$$a~=~\left( \begin{array}{cc} 2&-1\\-2&2 \end{array} \right),~~~~~
a^{-1}~=~\left( \begin{array}{cc} 1&1/2\\1&1 \end{array}
\right) \eqno (1) $$

\noindent
{}From it we have the relation between the simple roots ${\bf
r}_{j}$ and the fundamental dominant weight ${\bf \lambda}_{j}$:
$$\begin{array}{ll}
{\bf r}_{1}~=~2~{\bf \lambda}_{1}~-~2~{\bf \lambda}_{2},~~~~
&{\bf r}_{2}~=~-~{\bf \lambda}_{1}~+~2~{\bf \lambda}_{2} \\
{\bf \lambda}_{1}~=~{\bf r}_{1}~+~{\bf r}_{2},~~~~
&{\bf \lambda}_{2}~=~{\bf r}_{1}/2~+~{\bf r}_{2}
\end{array} \eqno (2) $$

An irreducible representation of $U_{q}B_{2}$ is denoted by its
highest weight ${\bf M}=(M_{1}~M_{2})$ and the states by ${\bf
m}=(m_{1}~m_{2})$:
$$\begin{array}{l}
{\bf M}~=~M_{1}~{\bf \lambda}_{1}~+~M_{2}~{\bf \lambda}_{2},~~~~
{\bf m}~=~m_{1}~{\bf \lambda}_{1}~+~m_{2}~{\bf \lambda}_{2}
\end{array} \eqno (3) $$

\noindent
The minimal representation is denoted by $(1~0)$, and the
adjoint representation by $(0~2)$. The Casimir $C_{2}({\bf M})$
is calculated by the following formula:
$$C_{2}({\bf M})~=~M_{1}^{2}~+~M_{1}M_{2}~+~M_{2}^{2}/2~
+~3M_{1}~+~2M_{2} \eqno (4)$$

Through a standard method [1] we draw the block weight diagrams
for the representations $(1~0)$ and $(0~2)$ in Fig.1.

\begin{center}
\setlength{\unitlength}{1.0mm}
\begin{picture}(94,111)(0,-30)
\put(11,60){\framebox(12,6){$1~0$}}
\put(11,50){\framebox(12,6){$\bar{1}~2$}}
\put(11,40){\framebox(12,6){$0~0$}}
\put(11,30){\framebox(12,6){$1~\bar{2}$}}
\put(11,20){\framebox(12,6){$\bar{1}~0$}}
\put(17,60){\line(0,-1){4}}
\put(16.5,50){\line(0,-1){4}}
\put(17.5,50){\line(0,-1){4}}
\put(17.5,40){\line(0,-1){4}}
\put(16.5,40){\line(0,-1){4}}
\put(17,30){\line(0,-1){4}}
\put(25,65){\makebox(0,0)[l]{$2$}}
\put(25,55){\makebox(0,0)[l]{$1$}}
\put(25,45){\makebox(0,0)[l]{$0$}}
\put(25,35){\makebox(0,0)[l]{$\bar{1}$}}
\put(25,25){\makebox(0,0)[l]{$\bar{2}$}}
\put(17,0){\makebox(0,0){a) Minimal representation}}

\put(71,70){\framebox(12,6){$0~2$}}
\put(71,60){\framebox(12,6){$1~0$}}
\put(71,20){\framebox(12,6){$\bar{1}~0$}}
\put(71,10){\framebox(12,6){$0~\bar{2}$}}
\put(60,50){\framebox(12,6){$\bar{1}~2$}}
\put(60,40){\framebox(12,6){$(0~0)_{2}$}}
\put(60,30){\framebox(12,6){$1~\bar{2}$}}
\put(82,50){\framebox(12,6){$2~\bar{2}$}}
\put(82,40){\framebox(12,6){$(0~0)_{0}$}}
\put(82,30){\framebox(12,6){$\bar{2},2$}}
\put(76.5,70){\line(0,-1){4}}
\put(77.5,70){\line(0,-1){4}}
\put(71,60){\line(0,-1){4}}
\put(82,60){\line(0,-1){4}}
\put(83,60){\line(0,-1){4}}
\put(71,50){\line(0,-1){4}}
\put(72,50){\line(0,-1){4}}
\put(83,50){\line(0,-1){4}}
\put(71,40){\line(0,-1){4}}
\put(72,40){\line(0,-1){4}}
\put(83,40){\line(0,-1){4}}
\put(71,30){\line(0,-1){4}}
\put(82,30){\line(0,-1){4}}
\put(83,30){\line(0,-1){4}}
\put(77.5,20){\line(0,-1){4}}
\put(76.5,20){\line(0,-1){4}}
\put(71,40){\line(3,-1){12}}
\put(83,50){\line(-3,-1){12}}
\put(85,75){\makebox(0,0)[l]{$4$}}
\put(85,65){\makebox(0,0)[l]{$3$}}
\put(96,55){\makebox(0,0)[l]{$2$}}
\put(96,45){\makebox(0,0)[l]{$0'$}}
\put(96,35){\makebox(0,0)[l]{$\bar{2}$}}
\put(58,55){\makebox(0,0)[r]{$1$}}
\put(58,45){\makebox(0,0)[r]{$0$}}
\put(58,35){\makebox(0,0)[r]{$\bar{1}$}}
\put(85,25){\makebox(0,0)[l]{$\bar{3}$}}
\put(85,15){\makebox(0,0)[l]{$\bar{4}$}}
\put(77,0){\makebox(0,0){b) Adjoint representation}}
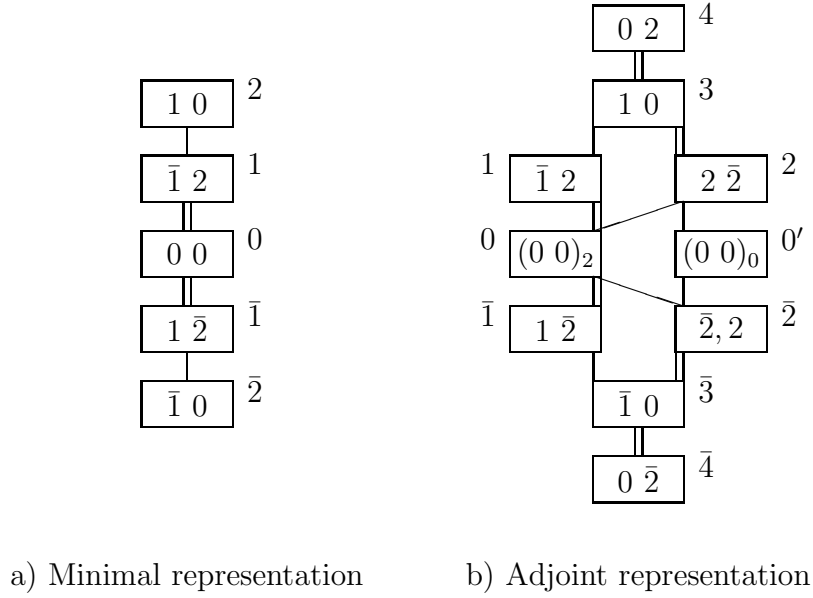
\put(50,-15){\makebox(0,0){{\bf Fig.1}. \parbox[t]{3.5in}{Block
weight diagrams for the minimal and\\ adjoint representations
of algebra $U_{q}B_{2}$}}}
\end{picture}
\end{center}

In order to simplify the notation we enumerate the states in
those two representations as shown near the blocks in Fig.1. In
terms of the enumerations for the states and the matrix bases
$E_{a~b}$:
$$\left(E_{a~b}\right)_{c~d}~=~\delta_{ac}~\delta_{bd} \eqno (5) $$

\noindent
we obtain the quantum representation matrices for two
representations as follows. For the minimal representation
$(1~0)$ we have:
$$\begin{array}{l}
D_{q}(e_{1})~=~\tilde{D}_{q}(f_{1})~=~E_{2~1}~+~E_{\bar{1}~
\bar{2}}\\
D_{q}(e_{2})~=~\tilde{D}_{q}(f_{2})~=~[2]^{1/2}~\left(~
E_{1~0}~+~E_{0~\bar{1}}~\right)\\
D_{q}(k_{1})~=~q~E_{2~2}~+~q^{-1}~E_{1~1}~+~E_{0~0}~+~q~E_{
\bar{1}~\bar{1}}~+~q^{-1}~E_{\bar{2}~\bar{2}} \\
D_{q}(k_{2})~=~E_{2~2}~+~q~E_{1~1}~+~E_{0~0}~+~q^{-1}~E_{
\bar{1}~\bar{1}}~+~E_{\bar{2}~\bar{2}} \end{array}
\eqno (6) $$

\noindent
and for the adjoint representation $(0~2)$ we have:
$$\begin{array}{rl}
D_{q}(e_{1})&=~\tilde{D}_{q}(f_{1})\\
&=~E_{3~1}~+~E_{2~0}~+~\displaystyle \left({[6] \over
[3][2]} \right)^{1/2} \left(E_{2~0'}~+~E_{0'~\bar{2}}\right)~
+~E_{0~\bar{2}}~+~E_{\bar{1}~\bar{3}}\\
D_{q}(e_{2})&=~\tilde{D}_{q}(f_{2})\\
&=~[2]^{1/2}~\left(~E_{4~3}~+~E_{3~2}~+~E_{1~0}~+~E_{0~\bar{1}}~
+~E_{\bar{2}~\bar{3}}~+~E_{\bar{3}~\bar{4}}~\right)\\
D_{q}(k_{1})&=~E_{4~4}~+~q~E_{3~3}~+~q^{2}~E_{2~2}~+~q^{-1}~
E_{1~1}~+~E_{0~0} \\
&~~~+~E_{0'~0'}~+~q~E_{\bar{1}~\bar{1}}~+~q^{-2}~E_{\bar{2}~
\bar{2}}~+~q^{-1}~E_{\bar{3}~\bar{3}}~+~E_{\bar{4}~\bar{4}} \\
D_{q}(k_{2})&=~q~E_{4~4}~+~E_{3~3}~+~q^{-1}~E_{2~2}~+~
q~E_{1~1}~+~E_{0~0} \\
&~~~+~E_{0'~0'}~+~q^{-1}~E_{\bar{1}~\bar{1}}~+~q~E_{\bar{2}~
\bar{2}}~+~E_{\bar{3}~\bar{3}}~+~q^{-1}~E_{\bar{4}~\bar{4}}
\end{array} \eqno (7) $$

\noindent
where, as usual, $[m]$ denotes:
$$[m]~=~{\displaystyle q^{m}~-~q^{-m} \over \displaystyle
q~-~q^{-1} } \eqno (8) $$

Since the negative lowest root ${\bf r}_{0}$ of $B_{2}$ is:
$${\bf r}_{0}~=~-~2{\bf \lambda}_{2}~=~-~{\bf r}_{1}~-~2{\bf
r}_{2} \eqno (9) $$

\noindent
the possible forms of the representation matrices of $e_{0}$ and
$f_{0}$, that correspond to ${\bf r}_{0}$, are as follows:
$$\begin{array}{l}
D_{q}(e_{0})~=~a_{1}E_{0~4}+a_{2}E_{0'~4}+a_{3}E_{\bar{1}~3}+
a_{4}E_{\bar{3}~1}+a_{5}E_{\bar{4}~0}+a_{6}E_{\bar{4}~0'} \\
D_{q}(f_{0})~=~b_{1}E_{4~0}+b_{2}E_{4~0'}+b_{3}E_{3~\bar{1}}+
b_{4}E_{1~\bar{3}}+b_{5}E_{0~\bar{4}}+b_{6}E_{0'~\bar{4}}
\end{array} \eqno (10) $$

{}From the quantum algebraic relations:
$$[~D_{q}(e_{0})~,~D_{q}(f_{j})~]~=~0,~~~~
[~D_{q}(f_{0})~,~D_{q}(e_{j})~]~=~0,~~~~j=1,~2 \eqno (11) $$

\noindent
we obtain
$$\begin{array}{l}
-~\left({\displaystyle [6] \over \displaystyle [3][2]}\right)^{
1/2} a_{2}~=~-~\left({\displaystyle [6] \over \displaystyle [3]
[2]}\right)^{1/2}a_{6}~=~a_{1}~=~a_{3}~=~a_{4}~=~a_{5} \\[2mm]
-~\left({\displaystyle [6] \over \displaystyle [3][2]}\right)^{
1/2} b_{2}~=~-~\left({\displaystyle [6] \over \displaystyle [3]
[2]}\right)^{1/2}
b_{6}~=~b_{1}~=~b_{3}~=~b_{4}~=~b_{5} \end{array} \eqno (12) $$

\noindent
It is easy to check that the quantum Serre relations are not
satisfied:
$$\begin{array}{l}
D_{q}(e_{0})^{2}D_{q}(e_{2})~-~\left([4]/[2]\right)D_{q}(e_{0})
D_{q}(e_{2})D_{q}(e_{0})~+~D_{q}(e_{2})D_{q}(e_{0})^{2}\\
=~-~\left(q^{-1}-q\right)^{2}\displaystyle \left( {[4][3] \over
[6] }\right)~a_{1}^{2}~[2]^{1/2}\left(E_{\bar{4}~3}~+~E_{\bar{3}~
4}\right)~\neq ~0 \\
D_{q}(f_{0})^{2}D_{q}(f_{2})~-~\left([4]/[2]\right)D_{q}(f_{0})
D_{q}(f_{2})D_{q}(f_{0})~+~D_{q}(f_{2})D_{q}(f_{0})^{2}\\
=~-~\left(q^{-1}-q\right)^{2}\displaystyle \left( {[4][3] \over
[6] }\right)~b_{1}^{2}~[2]^{1/2}\left(E_{3~\bar{4}}~+~E_{4~
\bar{3}}\right)~\neq ~0 \end{array} \eqno (13) $$

\noindent
The commutator of $D_{q}(e_{0})$ and $D_{q}(f_{0})$ does not
satisfy the quantum algebraic relations, either. Therefore, the
representation matrix $D_{q}(e_{0})$ does not exist for the
adjoint representation of $U_{q}B_{2}$.

\vspace{15mm}
\noindent
{\bf 3. Solutions for the Minimal Representation}

\vspace{3mm}
In the fusion formulas, the $\breve{R}_{q}^{{\rm adj}}(x)$
matrix for the adjoint representation is expressed in terms of
the $\breve{R}_{q}(x)$ matrix for the minimal representation.
In this section we compute the $\breve{R}_{q}(x)$ matrix for
the minimal representation firstly. As a matter of fact, the
$e_{0}$ matrix exists in the minimal representation of
$U_{q}B_{2}$ so that the corresponding solution $\breve{R}_{q}
(x)$ was computed [3,1] by the standard method based on
Jimbo's theorem.

The Clebsch-Gordan series for the direct product of two
minimal representations is as follows:
$$(1~0)~\otimes (1~0)~=~(2~0)~\oplus~(0~2)~\oplus~(0~0)
\eqno (14) $$

\noindent
Denote by ${\cal P}_{{\bf N}}$ the project operator that is the
product of two quantum Clebsch-Gordan coefficients [1]:
$${\cal P}_{{\bf N}}~=~\left(C_{q}\right)_{{\bf N}}~\left(
\tilde{C}_{q}\right)_{{\bf N}} \eqno (15) $$

\noindent
By making use of the standard method based on Jimbo's theorem,
we obtain the $\breve{R}_{q}'(x)$
matrix for the minimal representation as follows [3,1]:
$$\begin{array}{rl}
\breve{R}_{q}'(x)&=~(1-xq^{4})(1-xq^{6})~{\cal P}_{(2~0)} \\
&+~(x-q^{4})(1-xq^{6})~{\cal P}_{(0~2)} \\
&+~(x-q^{4})(x-q^{6})~{\cal P}_{(0~0)} \end{array} \eqno (16)$$

\noindent
where a prime is added on $\breve{R}_{q}'(x)$ in order to
distinguish it from the additional solution $\breve{R}_{q}(x)$
given in Eq.(17). In this form of $\breve{R}_{q}'(x)$, it cannot
be proportional to the projector operator ${\cal P}_{(0~2)}$
that maps the direct product space onto the space of the adjoint
representation. In the same paper [3] Jimbo pointed out that
there is another solution related to the algebra $U_{q}
A_{4}^{(2)}$:
$$\begin{array}{rl}
\breve{R}_{q}(x)&=~(1-xq^{4})(1+xq^{10})~{\cal P}_{(2~0)} \\
&+~(x-q^{4})(1+xq^{10})~{\cal P}_{(0~2)} \\
&+~(1-xq^{4})(x+q^{10})~{\cal P}_{(0~0)} \end{array} \eqno (17)
$$

\noindent
Now, we know [5,1] that because the Clebsch-Gordan series (14)
contains only three representations including an identity
representation $(0~0)$, we can obtain two independent solutions
of the Yang-Baxter equation given in Eqs.(16) and (17) in terms
of Yang-Baxterization. The solution (17), called non-standard
one, has a good property:
$$\breve{R}_{q}(q^{-4})~=~(q^{-4}-q^{4})(1+q^{6})~{\cal P}_{
(0~2)} \eqno (18) $$

\noindent
namely, $\breve{R}_{q}(q^{-4})$ is proportional to the project
operator ${\cal P}_{(0~2)}$ onto the adjoint representation:
$$\breve{R}_{q}(q^{-4})~\left(~V_{(1~0)}~\otimes~V_{(1~0)}~
\right)~=~V_{(0~2)} \eqno (19) $$

\noindent
It is the key point for computing the solution related to the
adjoint representation from Eq.(17).

Solution (17) is a $25 \times 25$ symmetric matrix on the direct
product space $V_{(1~0)}\otimes V_{(1~0)}$. The row (column)
indices are denoted by $m_{1}m_{2}$, where both $m_{1}$ and
$m_{2}$ take the values $2, ~1,~0,~\bar{1}$, and $\bar{2}$.
$\breve{R}_{q}(x)$ has the following symmetries:
$$\begin{array}{rl}
\breve{R}_{q}(x)_{m_{1}m_{2}~m_{3}m_{4}}
&=~\breve{R}_{q}(x)_{m_{3}m_{4}~m_{1}m_{2}} \\
&=~\breve{R}_{q}(x)_{\bar{m}_{2}\bar{m}_{1}~\bar{m}_{4}
\bar{m}_{3}} \\
&=~-x^{2}~q^{14}~\breve{R}_{q^{-1}}(x^{-1})_{m_{2}m_{1}~m_{4}
m_{3}} \\
\breve{R}_{q}(1)_{m_{1}m_{2}~m_{3}m_{4}}&=~(1-q^{4})(1+q^{10})~
\delta_{m_{1}m_{3}}~\delta_{m_{2}m_{4}} \end{array}
\eqno (20) $$

\noindent
where $\bar{0}=0$.

$\breve{R}_{q}(x)$ given in Eq.(17) satisfies the
weight conservation condition, namely, $\breve{R}_{q}(x)$ is a
block matrix with four $1\times 1$, eight $2\times 2$ and one
$5\times 5$ submatrices. Through straightforward calculation we
obtain the explicit form for $\breve{R}_{q}(x)$. Owing to the
symmetries (20) we only need to list the results as follows:

a) Four $1\times 1$ submatrices.
$$\breve{R}_{q}(x)_{22~22}~=~\breve{R}_{q}(x)_{11~11}~=~
(1-xq^{4})(1+xq^{10})  \eqno (21a) $$

b) Eight $2\times 2$ submatrices.
$$\begin{array}{rl}
\breve{R}_{q}(x)_{21~21}&=~\breve{R}_{q}(x)_{20~20}~=~
\breve{R}_{q}(x)_{2\bar{1}~2\bar{1}}~=~\breve{R}_{q}(x)_{10~10}
\\
&=~(1-q^{4})x(1+xq^{10})\\[2mm]
\breve{R}_{q}(x)_{21~12}&=~\breve{R}_{q}(x)_{20~02}~=~
\breve{R}_{q}(x)_{2\bar{1}~\bar{1}2}~=~\breve{R}_{q}(x)_{10~01}
\\
&=~q^{2}(1-x)(1+xq^{10}) \end{array} \eqno (21b) $$

c) One $5\times 5$ submatrix.
$$\begin{array}{rl}
\breve{R}_{q}(x)_{2\bar{2}~2\bar{2}}
&=~(1-q^{4}) x \left\{(1+q^{4})-xq^{4}(1-q^{6})\right\}\\
\breve{R}_{q}(x)_{1\bar{1}~1\bar{1}}
&=~(1-q^{4}) x \left\{(1+q^{8})-xq^{8}(1-q^{2})\right\}\\
\breve{R}_{q}(x)_{00~00}
&=~q^{2}(1-x)(1+xq^{10})+x(1-q^{4})(1+q^{10})\\
\breve{R}_{q}(x)_{2\bar{2}~1\bar{1}}
&=~-x(1-x)q^{6}(1-q^{4})\\
\breve{R}_{q}(x)_{2\bar{2}~00}
&=~x(1-x)q^{7}(1-q^{4})\\
\breve{R}_{q}(x)_{2\bar{2}~\bar{1}1}
&=~-x(1-x)q^{8}(1-q^{4})\\
\breve{R}_{q}(x)_{1\bar{1}~00}
&=~-x(1-x)q^{9}(1-q^{4}) \\
\breve{R}_{q}(x)_{2\bar{2}~\bar{2}2}&=~\breve{R}_{q}(x)_{1
\bar{1}~\bar{1}1}~=~q^{4}(1-x)(1+xq^{6})
 \end{array} \eqno (21c) $$

\vspace{15mm}
\noindent
{\bf 4. Fusion Formulas}

\vspace{3mm}
The project operator $\breve{R}_{q}(q^{-4})$ maps the direct
product space $V_{(1~0)}\otimes V_{(1~0)}$ of two minimal
representations onto the representation space $V_{(0~2)}$ of
the adjoint representation. The solution $\breve{R}_{q}^{{\rm
adj}}(x)$ of the Yang-Baxter equation related to the adjoint
representation of $U_{q}B_{2}$ is applied on the direct product
space $V_{(0~2)}\otimes V_{(0~2)}$:
$$V_{(0~2)}\otimes V_{(0~2)}~=~\left\{\breve{R}_{q}(q^{-4})
\otimes \breve{R}_{q}(q^{-4}) \right\}~\left\{V_{(1~0)}\otimes
V_{(1~0)}\otimes V_{(1~0)}\otimes V_{(1~0)}\otimes \right\}
\eqno (22) $$

\noindent
According to the fusion formulas, $\breve{R}_{q}^{{\rm adj}}(x)$
can be expressed as the following product [6,1]:
$$\breve{R}_{q}^{{\rm adj}}(x)~=~
\left({\bf 1}\otimes \breve{R}_{q}(xq^{4})\otimes {\bf 1}
\right)~\left(\breve{R}_{q}(x)\otimes \breve{R}_{q}(x)\right)~
\left({\bf 1}\otimes \breve{R}_{q}(xq^{-4})\otimes {\bf 1}
\right) \eqno (23) $$

Now, we are going to sketch the proof. First of all, we show
that $\breve{R}_{q}^{{\rm adj}}(x)$ given in Eq.(23) is a matrix
on the space (22). From the Yang-Baxter equation satisfied by
$\breve{R}_{q}(x)$:
$$\begin{array}{l}
\left({\bf 1}\otimes \breve{R}_{q}(x)\right)
\left(\breve{R}_{q}(xy)\otimes {\bf 1} \right)
\left({\bf 1}\otimes \breve{R}_{q}(y)\right) \\
=~\left(\breve{R}_{q}(y)\otimes {\bf 1} \right)
\left({\bf 1}\otimes \breve{R}_{q}(xy)\right)
\left(\breve{R}_{q}(x)\otimes {\bf 1} \right) \end{array} \eqno
(24) $$

\noindent
we have:
$$\begin{array}{l}
\breve{R}_{q}^{{\rm adj}}(x)~\left\{V_{(0~2)}\otimes V_{(0~2)}
\right\}\\
=~\left({\bf 1}\otimes \breve{R}_{q}(xq^{4})\otimes {\bf 1}
\right)~
\left({\bf 1}\otimes {\bf 1}\otimes \breve{R}_{q}(x)\right)~\\
{}~~~\cdot~\left(\breve{R}_{q}(x)\otimes {\bf 1}\otimes {\bf 1}
\right)~\left({\bf 1}\otimes \breve{R}_{q}(xq^{-4})\otimes
{\bf 1}\right) \left(\breve{R}_{q}(q^{-4}) \otimes {\bf 1}
\otimes {\bf 1} \right)\\
{}~~~\cdot~\left({\bf 1}\otimes {\bf 1} \otimes \breve{R}_{q}(
q^{-4}) \right)~\left\{V_{(1~0)}\otimes V_{(1~0)}\otimes
V_{(1~0)}\otimes V_{(1~0)} \right\}\\
=~\left({\bf 1}\otimes \breve{R}_{q}(xq^{4})\otimes {\bf 1}
\right)~\left({\bf 1}\otimes {\bf 1}\otimes \breve{R}_{q}(x)
\right)~\left({\bf 1}\otimes \breve{R}_{q}(q^{-4})\otimes
{\bf 1}\right)\\
{}~~~\cdot~\left(\breve{R}_{q}(xq^{-4})\otimes {\bf 1} \otimes
{\bf 1}\right) \left({\bf 1}\otimes \breve{R}_{q}(x) \otimes
{\bf 1} \right) ~\left({\bf 1}\otimes {\bf 1} \otimes
\breve{R}_{q}(q^{-4}) \right)\\
{}~~~\cdot~\left\{V_{(1~0)}\otimes V_{(1~0)}\otimes V_{(1~0)}
\otimes V_{(1~0)} \right\} \end{array} $$

$$\begin{array}{l}
=~\left({\bf 1}\otimes {\bf 1}\otimes \breve{R}_{q}(q^{-4})
\right)~\left({\bf 1}\otimes \breve{R}_{q}(x)\otimes {\bf 1}
\right)~\left(\breve{R}_{q}(xq^{-4})\otimes {\bf 1} \otimes
{\bf 1}\right)\\
{}~~~\cdot~\left({\bf 1}\otimes {\bf 1}\otimes \breve{R}_{q}(
xq^{4})\right)~\left({\bf 1}\otimes \breve{R}_{q}(x) \otimes
{\bf 1} \right) ~\left({\bf 1}\otimes {\bf 1} \otimes
\breve{R}_{q}(q^{-4}) \right)\\
{}~~~\cdot~\left\{V_{(1~0)}\otimes V_{(1~0)}\otimes V_{(1~0)}
\otimes V_{(1~0)} \right\}\\
=~\left({\bf 1}\otimes {\bf 1}\otimes \breve{R}_{q}(q^{-4})
\right)~\left({\bf 1}\otimes \breve{R}_{q}(x)\otimes {\bf 1}
\right)~\left(\breve{R}_{q}(xq^{-4})\otimes {\bf 1} \otimes
{\bf 1}\right)\\
{}~~~\cdot~\left({\bf 1}\otimes \breve{R}_{q}(q^{-4})\otimes {\bf
1}\right)~\left({\bf 1}\otimes {\bf 1}\otimes \breve{R}_{q}(x)
\right) ~\left({\bf 1}\otimes \breve{R}_{q}(xq^{4}) \otimes
{\bf 1}\right)\\
{}~~~\cdot~\left\{V_{(1~0)}\otimes V_{(1~0)}\otimes V_{(1~0)}
\otimes V_{(1~0)}\right\}\\
=~\left({\bf 1}\otimes {\bf 1}\otimes \breve{R}_{q}(q^{-4})
\right)~\left(\breve{R}_{q}(q^{-4})\otimes {\bf 1} \otimes {\bf
1} \right)~\left({\bf 1} \otimes \breve{R}_{q}(xq^{-4}) \otimes
{\bf 1}\right)\\
{}~~~\cdot~\left(\breve{R}_{q}(x) \otimes {\bf 1} \otimes {\bf 1}
\right)~ \left({\bf 1}\otimes {\bf 1}\otimes \breve{R}_{q}(x)
\right) ~\left({\bf 1}\otimes \breve{R}_{q}(xq^{4}) \otimes
{\bf 1}\right)\\
{}~~~\cdot~\left\{V_{(1~0)}\otimes V_{(1~0)}\otimes V_{(1~0)}
\otimes V_{(1~0)} \right\}\\
\subset~\left(\breve{R}_{q}(q^{-4})\otimes \breve{R}_{q}(q^{-4})
\right)~\left\{V_{(1~0)}\otimes V_{(1~0)}\otimes V_{(1~0)}
\otimes V_{(1~0)} \right\}\\
=~V_{(0~2)}\otimes V_{(0~2)} \end{array} $$

By making use of Eq.(24) successively, it is straightforward to
prove that $\breve{R}_{q}^{{\rm adj}}(x)$ satisfies
the Yang-Baxter equation, that is an equation on the direct
product space $V_{(1~0)}^{\otimes 6}$:
$$\begin{array}{l}
\left({\bf 1}\otimes {\bf 1}\otimes {\bf 1}\otimes
\breve{R}_{q}(xq^{4})\otimes {\bf 1}\right)
\left({\bf 1}\otimes {\bf 1}\otimes \breve{R}_{q}(x)\otimes
\breve{R}_{q}(x)\right)
\left({\bf 1}\otimes {\bf 1}\otimes {\bf 1}\otimes
\breve{R}_{q}(xq^{-4})\otimes {\bf 1} \right) \\
\cdot \left({\bf 1}\otimes \breve{R}_{q}(xyq^{4})\otimes {\bf 1}
\otimes {\bf 1}\otimes {\bf 1}\right)
\left(\breve{R}_{q}(xy)\otimes \breve{R}_{q}(xy)\otimes {\bf 1}
\otimes {\bf 1}\right)
\left({\bf 1}\otimes \breve{R}_{q}(xyq^{-4})\otimes {\bf 1}
\otimes {\bf 1}\otimes {\bf 1}\right) \\
\cdot \left({\bf 1}\otimes {\bf 1}\otimes {\bf 1}\otimes
\breve{R}_{q}(yq^{4})\otimes {\bf 1}\right)
\left({\bf 1}\otimes {\bf 1}\otimes {\bf 1}\otimes
\breve{R}_{q}(y)\otimes \breve{R}_{q}(y)\right)
\left({\bf 1}\otimes {\bf 1}\otimes {\bf 1}\otimes
\breve{R}_{q}(yq^{-4})\otimes {\bf 1}\right) \\[2mm]
=~\left({\bf 1}\otimes \breve{R}_{q}(yq^{4})\otimes {\bf 1}
\otimes {\bf 1}\otimes {\bf 1}\right)
\left(\breve{R}_{q}(y)\otimes \breve{R}_{q}(y)\otimes {\bf 1}
\otimes {\bf 1}\right)
\left({\bf 1}\otimes \breve{R}_{q}(yq^{-4})\otimes {\bf 1}
\otimes {\bf 1}\otimes {\bf 1}\right) \\
\cdot \left({\bf 1}\otimes {\bf 1}\otimes {\bf 1}\otimes
\breve{R}_{q}(xyq^{4})\otimes {\bf 1}\right)
\left({\bf 1}\otimes {\bf 1}\otimes
\breve{R}_{q}(xy)\otimes \breve{R}_{q}(xy)\right)
\left({\bf 1}\otimes {\bf 1}\otimes {\bf 1}\otimes
\breve{R}_{q}(xyq^{-4})\otimes {\bf 1}\right) \\
\cdot \left({\bf 1}\otimes \breve{R}_{q}(xq^{4})\otimes {\bf 1}
\otimes {\bf 1}\otimes {\bf 1}\right)
\left(\breve{R}_{q}(x)\otimes \breve{R}_{q}(x)\otimes {\bf 1}
\otimes {\bf 1}\right)
\left({\bf 1}\otimes \breve{R}_{q}(xq^{-4})\otimes {\bf 1}
\otimes {\bf 1}\otimes {\bf 1}\right)
\end{array} \eqno (25) $$

\vspace{15mm}
\noindent
{\bf 5. Explicit Form of the Solution for the Adjoint
Representation}

\vspace{3mm}
The Clebsch-Gordan series for the direct product of two adjoint
representations of $B_{2}$ is:
$$(0~2)~\otimes ~(0~2)~=~(0~4)~\oplus~(1~2)~\oplus~(2~0)~\oplus~
(0~2)~\oplus~(1~0)~\oplus~(0~0) \eqno (26) $$

\noindent
Both the solution $\breve{R}_{q}^{{\rm adj}}$ of the simple
Yang-Baxter equation and the solution $\breve{R}_{q}^{{\rm
adj}}(x)$ of the Yang-Baxter equation, related to the adjoint
representation of $U_{q}B_{2}$, can be
expanded by the project operators as follows:
$$\begin{array}{rl}
\breve{R}_{q}^{{\rm adj}}&=~{\cal P}_{(0~4)}~-~
q^{4}~{\cal P}_{(1~2)}~+~q^{6}~{\cal P}_{(2~0)}~-~q^{10}~{\cal
P}_{(0~2)}~+~q^{12}~{\cal P}_{(1~0)}~+~q^{16}~{\cal P}_{(0~0)}
\end{array} \eqno (27) $$
$$\begin{array}{rl}
\breve{R}_{q}^{{\rm adj}}(x)&=~\Lambda_{(0~4)}(x,q)~{\cal P}_{
(0~4)}~+~
\Lambda_{(1~2)}(x,q)~{\cal P}_{(1~2)}~+~\Lambda_{(2~0)}(x,q)~
{\cal P}_{(2~0)}\\
&~~~+~\Lambda_{(0~2)}(x,q)~{\cal P}_{(0~2)}~+~\Lambda_{(1~0)}(
x,q)~{\cal P}_{(1~0)}~+~\Lambda_{(0~0)}(x,q)~{\cal P}_{(0~0)}
\end{array} \eqno (28) $$
$$\breve{R}_{q}^{{\rm adj}}(0)~=~\breve{R}_{q}^{{\rm adj}}
\eqno (29) $$

\noindent
where, as usual, the project operators are the product of two
quantum Clebsch-Gordan matrices:
$${\cal P}_{{\bf N}}~=~\left(C_{q}^{(0~2)(0~2)}\right)_{{\bf N}}
{}~\left(\tilde{C}_{q}^{(0~2)(0~2)}\right)_{{\bf N}} \eqno (30) $$

\noindent
Now, we are going to compute the coefficients $\Lambda_{{\bf N}}
(x,q)$:
$$\breve{R}_{q}^{{\rm adj}}(x)~|~{\bf N}~,~{\bf N}~\rangle~=~
\Lambda_{{\bf N}}(x,q)~|~{\bf N}~,~{\bf N}~\rangle \eqno (31) $$

\noindent
In the computation, we need the quantum Clebsch-Gordan
coefficients to combine the states $|m_{1},m_{2},m_{3},m_{4}
\rangle \equiv |m_{1}\rangle |m_{2}\rangle |m_{3}\rangle
|m_{4}\rangle$ in the space
$V_{(1~0)}\otimes V_{(1~0)}\otimes V_{(1~0)}\otimes V_{(1~0)}$
into the state $|{\bf N},{\bf N}\rangle$.

Firstly, through the standard calculation, we obtain the
quantum Clebsch-Gordan coefficients for the adjoint
representation of $U_{q}B_{2}$. Denote by $|~(0~2)~,~m~\rangle$
the states in the adjoint representation, and by $|~m_{1}~m_{2}~
\rangle \equiv |m_{1}\rangle |m_{2}\rangle$ the states
in the space $V_{(1~0)}\otimes V_{(1~0)}$, where the states is
described by the enumerations given in Fig.1. Owing to the
symmetry of the quantum Clebsch-Gordan coefficients:
$$|~(0~2)~,~m~\rangle~=~\displaystyle \sum_{m_{1}m_{2}}~\left(
C_{q}\right)_{m_{1}m_{2}(0~2)m}~|~m_{1}~m_{2}~ \rangle $$
$$\begin{array}{rl}
\left(C_{q}\right)_{m_{1}m_{2}(0~2)m}&=~
-~\left(C_{q^{-1}}\right)_{m_{2}m_{1}(0~2)m}\\
&=~-~\left(C_{q^{-1}}\right)_{\bar{m}_{1}\bar{m}_{2}(0~2)
\bar{m}} \end{array} \eqno (32) $$

\noindent
we only need to list the following Clebsch-Gordan coefficients:
$$\begin{array}{rl}
|~(0~2)~,~4~\rangle&=~\left([2]/[4]\right)^{1/2}~\left\{~q^{-1}~
| ~2~ 1~ \rangle~-~q~| ~1~ 2~ \rangle ~\right\} \\
|~(0~2)~,~3~\rangle&=~[2]^{-1/2}~f_{2}~|~(0~2)~,~4~\rangle~=~
\left([2]/[4]\right)^{1/2}~\left\{~q^{-1}~
| ~2~ 0~ \rangle~-~q~| ~0~2~ \rangle ~\right\} \\
|~(0~2)~,~2~\rangle&=~[2]^{-1/2}~f_{2}~|~(0~2)~,~3~\rangle~=~
\left([2]/[4]\right)^{1/2}~\left\{~q^{-1}~|~2~\bar{1}~\rangle~
-~q~|~\bar{1}~ 2~ \rangle ~\right\} \\
|~(0~2)~,~1~\rangle&=~f_{1}~|~(0~2)~,~3~\rangle~=~
\left([2]/[4]\right)^{1/2}~\left\{~q^{-1}~
| ~1~ 0~ \rangle~-~q~| ~0~ 1~ \rangle ~\right\} \\
|~(0~2)~,~0~\rangle&=~[2]^{-1/2}~f_{2}~|~(0~2)~,~1~\rangle \\
&=~\left([2]/[4]\right)^{1/2}~\left\{~| ~1~ \bar{1}~ \rangle~
+~(q^{-1}-q)~|~0~0~\rangle~
-~|~\bar{1}~ 1~ \rangle ~\right\} \\
|~(0~2)~,~0'~\rangle&=~\left([3][2]/[6]\right)^{1/2}~\left\{~
f_{1}~|~(0~2)~,~2~\rangle ~-~|~(0~2)~,~0~\rangle ~\right\}\\
&=~[2]\left([3]/[6][4]\right)^{1/2}~\left\{~|~2~ \bar{2}~
\rangle~+~(q^{-2}-1)~|~1~ \bar{1}~ \rangle \right. \\
&~~\left.~-~(q^{-1}-q)~|~0~0~\rangle~+~(1-q^{2})~|~\bar{1}~1~
\rangle~-~|~\bar{2}~2~\rangle ~\right\}
\end{array} \eqno (33) $$

{}From Eq.(33) we are able to compute the expansive expressions
for the highest weight states in the Clebsch-Gordan series (26):
$$\begin{array}{rl}
|~(0~4)~,~(0~4)~\rangle &=~|~(0~2)~,~4~\rangle |~(0~2)~,~4~
\rangle \\
&=\left([2]/[4]\right)~\left\{~q^{-2}~|~2~1~2~1~\rangle~-~|~
2~1~1~2~\rangle \right. \\
&~~\left.~-~|~1~2~2~1~\rangle~+~q^{2}~|~1~2~1~2~\rangle~\right\}
\end{array} \eqno (34a) $$
$$\begin{array}{rl}
|~(1~2)&,~(1~2)~\rangle \\
&=~\left([2]/[4]\right)^{1/2}~\left\{~
q^{-1}~|~(0~2)~,~4~\rangle |~(0~2)~,~3~\rangle~-~
q~|~(0~2)~,~3~\rangle |~(0~2)~,~4~\rangle \right\}\\
&=\left([2]/[4]\right)^{3/2}~\left\{~q^{-3}~|~2~1~2~0~\rangle~-~
q^{-1}~|~2~1~0~2~\rangle~-~q^{-1}~|~1~2~2~0~\rangle  \right. \\
&~~~+~q~|~1~2~0~2~\rangle~
-~q^{-1}~|~2~0~2~1~\rangle~+~q~|~0~2~2~1~\rangle \\
&~~\left.~+~q~|~2~0~1~2~\rangle~-~q^{3}~|~0~2~1~2~\rangle~
\right\} \end{array} \eqno (34b) $$

\noindent
where we see that the second half terms of Eqs.(34a) and (34b)
can be obtained from the first half terms by exchanging:
$$F(q)~|~m_{1}~m_{2}~m_{3}~m_{4}~\rangle~\longrightarrow~
\pm~F(q^{-1})~|~m_{4}~m_{3}~m_{2}~m_{1}~\rangle \eqno (35) $$

\noindent
where the plus sign stands for Eq.(34a), and the minus sign for
Eq.(34b). In the following we will use the abbreviatory notation
$({\rm S~~terms})$ (for eq.(35) with plus sign) or $({\rm A~~
terms})$ (minus sign) to replace the second half terms,
respectively. In this way equations (34a) and (34b) are
rewritten as follows:
$$\begin{array}{rl}
|~(0~4)~,~(0~4)~\rangle &=\left([2]/[4]\right)~\left\{~
q^{-2}~|~2~1~2~1~\rangle~-~{1 \over 2}~|~2~1~1~2~
\rangle \right. \\
&~~\left.~-~{1 \over 2}~|~1~2~2~1~\rangle~+~({\rm S~~terms})~
\right\} \end{array}  $$
$$\begin{array}{rl}
|~(1~2)&,~(1~2)~\rangle ~
=\left([2]/[4]\right)^{3/2}~\left\{~q^{-3}~|~2~1~2~0~\rangle~
-~q^{-1}~|~2~1~0~2~\rangle \right. \\
&~~\left.~-~q^{-1}~|~1~2~2~0~\rangle ~+~q~|~1~2~0~2~\rangle~
+~({\rm A~~terms})~ \right\} \end{array} $$

\noindent
In the same way we have:
$$\begin{array}{rl}
|~(2~0)&,~(2~0)~\rangle \\
&=~\left([3]\right)^{-1/2}~\left\{~
q^{-1}~|~(0~2)~,~4~\rangle |~(0~2)~,~2~\rangle~-~
|~(0~2)~,~3~\rangle |~(0~2)~,~3~\rangle \right. \\
&~~\left.~+~
q~|~(0~2)~,~2~\rangle |~(0~2)~,~4~\rangle \right\}\\
&=\left([2]/[4]\right)[3]^{-1/2}~\left\{~q^{-3}~|~2~1~2~\bar{1}~
\rangle~-~q^{-1}~|~2~1~\bar{1}~2~\rangle ~-~q^{-1}~|~1~2~2~
\bar{1}~\rangle \right. \\
&~~~+~q~|~1~2~\bar{1}~2~\rangle ~-~q^{-2}~|~2~0~2~0~\rangle ~
+~{1 \over 2}~|~2~0~0~2~\rangle \\
&~~\left.~+~{1 \over 2}~|~0~2~2~0~\rangle~+~({\rm S~~terms})
\right\} \end{array} \eqno (34c) $$
$$\begin{array}{rl}
|~(0~2)&,~(0~2)~\rangle \\
&=~[3]^{-1}\left([6][5][2]/[10][4]\right)^{1/2}~
\left\{~q^{-3}~|~(0~2)~,~4~\rangle |~(0~2)~,~0~\rangle \right.\\
&~~~-~q^{-3}\left([3][2]/[6] \right)^{1/2}~|~(0~2)~,~4~\rangle
|~(0~2)~,~0'~\rangle ~-~q^{-1}~|~(0~2)~,~3~\rangle |~(0~2)~,~ 1~
\rangle \\
&~~~+~ q~|~(0~2)~,~1~\rangle |~(0~2)~,~ 3~\rangle ~
+~q^{3}\left([3][2]/[6]\right)^{1/2}~|~(0~2)~,~0'~\rangle
|~(0~2)~,~4~\rangle \\
&~~\left.~-~q^{3}~|~(0~2)~,~0~\rangle |~(0~2)~,~4~\rangle ~
\right\} \\
&=\left([2]^{2}/[4]\right)\left([5][2]/[10][6][4]\right)^{1/2}~
\left\{~-~q^{-4}~|~2~1~2~\bar{2}~\rangle~+~q^{-2}~|~2~1~1~
\bar{1}~\rangle~\right.\\
&~~~-~q^{-6}~|~2~1~\bar{1}~1~\rangle ~
+~q^{-4}~|~2~1~\bar{2}~2~\rangle ~
+~q^{-2}~|~1~2~2~\bar{2}~\rangle ~-~|~1~2~1~\bar{1}~\rangle \\
&~~~+~q^{-4}~|~1~2~\bar{1}~1~\rangle~-~q^{-2}~|~1~2~\bar{2}~2~
\rangle ~+~(q^{-1}-q)\left([4]/[2]\right)~\left(~q^{-4}~|~2~1~0 ~
0~\rangle \right.\\
&~~\left.~-~q^{-2}~|~1~2~0~0~\rangle~\right)~+~\left([6]/[3][2]
\right)\left(~-~q^{-3}~|~2~0~1~0~\rangle ~+~q^{-1}~|~2~0~0~1~
\rangle \right.\\
&~~\left. \left.~+~q^{-1}~|~0~2~1~0~\rangle ~
-~q~|~0~2~0~1~\rangle \right)~+~({\rm A~~terms})~ \right\}
\end{array} \eqno (34d) $$
$$\begin{array}{rl}
|~(1~0)&,~(1~0)~\rangle \\
&=~\left([4]/[8][3]\right)^{1/2}~
\left\{~q^{-3}~|~(0~2)~,~4~\rangle |~(0~2)~,~\bar{1}~\rangle~-~
q^{-2}~|~(0~2)~,~3~\rangle |~(0~2)~,~0~\rangle \right. \\
&~~~+~ q^{-1}~|~(0~2)~,~2~\rangle |~(0~2)~,~ 1~\rangle ~
+~ q~|~(0~2)~,~1~\rangle |~(0~2)~,~ 2~\rangle \\
&~~\left.~-~q^{2}~|~(0~2)~,~0~\rangle |~(0~2)~,~3~\rangle ~
+~q^{3}~|~(0~2)~,~\bar{1}~\rangle |~(0~2)~,~4~\rangle ~
\right\} \\
&=[2]\left([8][4][3]\right)^{-1/2}~\left\{~
q^{-5}~|~2~1~0~\bar{1}~\rangle~-~q^{-3}~|~2~1~\bar{1}~0~
\rangle~-~q^{-3}~|~1~2~0~\bar{1}~\rangle \right. \\
&~~~+~q^{-1}~|~1~2~\bar{1}~0~\rangle ~
-~q^{-3}~|~2~0~1~\bar{1}~\rangle~-~\left(q^{-4}-q^{-2}\right)~
|~2~0~0~0~\rangle~+~q^{-3}~|~2~0~\bar{1}~1~\rangle \\
&~~~+~q^{-1}~|~0~2~1~\bar{1}~\rangle ~+~(q^{-2}-1)~|~0~2~0~0~
\rangle ~-~q^{-1}~|~0~2~\bar{1}~1~\rangle~+~q^{-3}~|~2~\bar{1}~
1~0~\rangle\\
&~\left.~\left.~-~q^{-1}~|~2~\bar{1}~0~1~\rangle ~
-~q^{-1}~|~\bar{1}~2~1~0~\rangle ~
+~q~|~\bar{1}~2~0~1~\rangle \right)~+~({\rm S~~terms}) \right\}
\end{array} \eqno (34e) $$
$$\begin{array}{rl}
|~(0~0)&,~(0~0)~\rangle \\
&=~\left([4]/[8][5]\right)^{1/2}~
\left\{~q^{-4}~|~(0~2)~,~4~\rangle |~(0~2)~,~\bar{4}~\rangle~-~
q^{-3}~
|~(0~2)~,~3~\rangle |~(0~2)~,~\bar{3}~\rangle \right. \\
&~~~+~ q^{-2}~|~(0~2)~,~2~\rangle |~(0~2)~,~\bar{2}~\rangle ~
+~ q^{-1}~|~(0~2)~,~1~\rangle |~(0~2)~,~\bar{1}~\rangle \\
&~~~-~|~(0~2)~,~0~\rangle |~(0~2)~,~0~\rangle ~
-~|~(0~2)~,~0'~\rangle |~(0~2)~,~0'~\rangle \\
&~~~+~q~|~(0~2)~,~\bar{1}~\rangle |~(0~2)~,~1~\rangle ~
+~ q^{2}~|~(0~2)~,~\bar{2}~\rangle |~(0~2)~,~2~\rangle \\
&~~\left.~-~q^{3}~|~(0~2)~,~\bar{3}~\rangle |~(0~2)~,~3~\rangle ~
+~ q^{4}~|~(0~2)~,~\bar{4}~\rangle |~(0~2)~,~4~\rangle \right\}
\\
&=[2]\left([8][5][4]\right)^{-1/2}~\left\{~
q^{-6}~|~2~1~\bar{1}~\bar{2}~\rangle~-~q^{-4}~|~2~1~\bar{2}~
\bar{1}~\rangle~
-~q^{-4}~|~1~2~\bar{1}~\bar{2}~\rangle \right. \\
&~~~+~q^{-2}~|~1~2~\bar{2}~\bar{1}~\rangle ~
-~q^{-5}~|~2~0~0~\bar{2}~\rangle~+~q^{-3}~|~2~0~\bar{2}~0~
\rangle~+~q^{-3}~|~0~2~0~\bar{2}~\rangle \\
&~~~-~q^{-1}~|~0~2~\bar{2}~0~\rangle ~
+~q^{-4}~|~2~\bar{1}~1~\bar{2}~\rangle~-~q^{-2}~|~2~\bar{1}~
\bar{2}~1~\rangle~-~q^{-2}~|~\bar{1}~2~1~\bar{2}~\rangle \\
&~~~+~|~\bar{1}~2~\bar{2}~1~\rangle ~
 +~q^{-3}~|~1~0~0~\bar{1}~\rangle ~-~q^{-1}~|~1~0~\bar{1}~0~
\rangle~-~q^{-1}~|~0~1~0~\bar{1}~\rangle \\
&~~~+~q~|~0~1~\bar{1}~0~\rangle ~
+~\left([3][2]/2[6]\right)\left(-2~|~2~\bar{2}~2~\bar{2}~
\rangle~-2~\left(q^{-4}-q^{-2}+q^{2}\right)~|~1~\bar{1}~1~
\bar{1}~\rangle \right. \\
&~~\left.~+~|~1~\bar{1}~\bar{1}~1~\rangle ~+~|~\bar{1}~1~1~
\bar{1}~\rangle ~+~|~2~\bar{2}~\bar{2}~2~\rangle ~
+~|~\bar{2}~2~2~\bar{2}~\rangle \right)\\
&~~~+~\left(q^{-1}-q\right)\left(~|~2~\bar{2}~0~0~\rangle ~
+~|~0~0~2~\bar{2}~\rangle ~-~q^{-1}~|~2~\bar{2}~1~\bar{1}~
\rangle -q^{-1}~|~1~\bar{1}~2~\bar{2}~\rangle \right. \\
&~~\left.~ -q^{2}~|~1~\bar{1}~0~0~\rangle ~-q^{2}~|~0~0~1~
\bar{1}~\rangle ~-~q~|~2~\bar{2}~\bar{1}~1~\rangle
-~q~|~\bar{1}~1~2~\bar{2}~\rangle \right) \\
&~~\left.~-~\left(q^{-1}-q\right)^{2}\left([4]/2[2]\right)~|~0~
0~0~0~\rangle~+~({\rm S~~terms}) \right\} \end{array} \eqno
(34f) $$

Now, substituting Eqs.(21), (23)) and (34) into Eq.(31), we
obtain $\Lambda_{{\bf N}}(x,q)$, and then, the solution
$\breve{R}_{q}^{{\rm adj}}(x)$ of the Yang-Baxter equation
related to the adjoint representation of $U_{q}B_{2}$ as
follows:
$$\begin{array}{rl}
\breve{R}_{q}^{{\rm adj}}(x)&=~(q^{4}-x)(1-x)(1+xq^{10})(1+
xq^{14}) \\
&~~~\cdot~\left\{ (1-xq^{4})(1+xq^{6})(1-xq^{8})(1+xq^{10})~
{\cal P}_{(0~4)} \right.\\
&~~~+~(x-q^{4})(1+xq^{6})(1-xq^{8})(1+xq^{10})~{\cal P}_{(1~2)}
\\
&~~~+~(1-xq^{4})(x+q^{6})(1-xq^{8})(1+xq^{10})~{\cal P}_{(2~0)}
\\
&~~~+~(x-q^{4})(x+q^{6})(1-xq^{8})(1+xq^{10})~{\cal P}_{(0~2)}\\
&~~~+~(x-q^{4})(1+xq^{6})(x-q^{8})(1+xq^{10})~{\cal P}_{(1~0)}\\
&~~\left.~+~(1-xq^{4})(x+q^{6})(1-xq^{8})(x+q^{10})~{\cal P}_{
(0~0)}\right\} \end{array} \eqno (36) $$

\noindent
where the common factor $(q^{4}-x)(1-x)(1+xq^{10})(1+xq^{14})$
can be removed. In principle, this method can be generalized to
the solutions of the Yang-Baxter equation related to the adjoint
representations of $U_{q}B_{\ell}$, $U_{q}C_{\ell}$ and $U_{q}
D_{\ell}$.

\vspace{15mm}
\noindent
{\bf 6. Rational Solution for the Adjoint Representation}

Through a standard limit process [1] we obtain the corresponding
rational solution $R^{{\rm adj}}(u/\eta)$ for the adjoint
representation of $U_{q}B_{2}$:
$$\begin{array}{rl}
R^{{\rm adj}}(u/\eta)&=~\displaystyle \lim_{q\rightarrow 1}~
{\displaystyle P~\breve{R}_{q}^{{\rm adj}}(q^{2u/\eta}) \over
\displaystyle \left( 1~-~q^{2u/\eta}\right)^{2} }\\
&=~4~\left\{~(1+2u/\eta)(1+4u/\eta)~\left(~P_{(0~4)}~+~P_{(2~
0)}~+~P_{(0~0)}~ \right) \right. \\
&~~~~+~(1-2u/\eta)(1+4u/\eta)~\left(~P_{(1~2)}~+~P_{(0~2)}~
 \right)\\
&~~\left.~~+~(1-2u/\eta)(1-4u/\eta)~P_{(1~0)}
\right\} \end{array} \eqno (37) $$

\noindent
where $P$ is the transposition operator, and
$$P_{{\bf N}}~=~\displaystyle \lim_{q \rightarrow 1}~{\cal
P}_{{\bf N}} $$

\vspace{2.0cm}
{\bf Acknowledgment}. This paper was supported by the National Natural
Science Foundation of China and Grant No. LWTZ-1298 of Chinese Academy
of Sciences.


\newpage
\begin{thebibliography}{99}
\bibitem{1}Zhong-Qi Ma, {\it Yang-Baxter Equation and Quantum
Enveloping Algebras}, World Scientific, Singapore, 1993.

\bibitem{2}C. N. Yang, {\it Phys. Rev. Lett}., {\bf
19}(1967)1312; R. J. Baxter, {\it Ann. Phys}., {\bf
70}(1972)193.

\bibitem{3}M. Jimbo, {\it Commun. Math. Phys}.,{\bf
102}(1986)537.

\bibitem{4}Zhong-Qi Ma, {\it Commun. Theor. Phys}., {\bf
15}(1991)37.

\bibitem{5}V. F. R. Jones, {\it Commun. Math. Phys}. {\bf
125}(1989)459;
Y. Cheng, M. L. Ge and K. Xue, {\it Commun. Math. Phys}. {\bf
136}(1991)195.

\bibitem{6} M. Jimbo, {\it Introduction to the Yang-Baxter
Equation}, in "{\it Braid Group, Knot Theory and Statistical
Mechanics}", Ed. by C. N. Yang and M. L. Ge, World Scientific,
Singapore, 1989, P.111.


\end{thebibliography}
\end{document}